# The Projectile inside the Loop

**Gabriele U. Varieschi,** Loyola Marymount University, Los Angeles, CA

The *loop-the-loop* demonstration[1] is one of the favorite toys used in introductory physics courses. In this simple device a small sphere typically rolls down an incline and then continues around a circular track, which constitutes the "loop." By using the principle of conservation of mechanical energy, students are usually asked to find the initial conditions that enable the moving body to "safely" make it around the loop.[2] At times, I tried to ask my students an unusual question: "What happens to the body if it doesn't go around the loop and falls inside it?" In this paper I will detail the answer to this question and describe a simple experimental activity related to this interesting problem.

The student answers are usually contradictory. Some suggest that the object would simply fall straight down, after loosing contact with the rail of the loop. Others suppose that the body would continue to follow a somewhat circular trajectory in the air, with a different radius of curvature, compared to the loop radius. In general, many students fail to recognize that the object, once the contact with the rail is lost, simply follows the kinematical rules of a projectile motion.

**The Description of the Problem**

Consider the loop-the-loop apparatus, schematically described in Figure 1, composed of an initial "ramp" and a circular track of radius R. In the actual demonstration a spherical object (represented by the red dot in the figure) of mass m, radius r and moment of inertia $I = \frac{2}{5}mr^2$, starts moving from rest at a certain initial height, and then rolls down the ramp and inside the loop. We can assume a pure rolling motion without slipping, and neglect air resistance or any other energy loss in the motion.[3] The red dot in the figure should actually represent the position of the center of mass of the rolling sphere, for which all the following analysis will apply. Alternatively, we can assume the radius r of the sphere being much smaller than the track radius R, so that a point-like object, as in Fig. 1, can represent the body.

It is a well-known result that the minimum initial height required for the ball to make the loop is exactly $\frac{27}{10}R$, as measured from the bottom level of the loop. If the ball is released from an initial height $h_i$ less than 2.7 R, it will not gain enough speed to complete the loop. At this point two options are still open: if the release height is in the range $0 \le h_i \le R$, the ball will simply rise to the same height inside the loop, and then keep oscillating back and forth, without ever losing contact with the track. The second option is the most interesting and related to our original problem. If the initial height is in the range $R < h_i < 2.7R$ (indicated by the green bar on the y axis in Fig. 1), the ball will reach a position (denoted by point P and related angle θ, in Fig. 1) in the upper left quadrant of the loop, at which it will loose contact with the rail, while still possessing a non-zero speed, $v_P$.

The contact with the track is lost at the point where the normal force **N**, of the track on the ball becomes zero, i.e., when the radial component of the weight alone will provide the centripetal force, $mg \sin\theta = m\frac{v_P^2}{R}$, so that:

$$v_P^2 = gR\sin\theta, \tag{1}$$

at position P in the figure. Imposing standard conservation of mechanical energy, between initial position at height $h_i$ and final position at height $h_f = R(1+\sin\theta)$, with linear and angular velocities denoted by $v_P$ and $\omega_P = v_P/r$ respectively, we can write:

$$mgh_i = mgh_f + \tfrac{1}{2}mv_P^2 + \tfrac{1}{2}I\omega_P^2. \tag{2}$$

If we express the initial height as $h_i = aR$, where $a$ is a number in the range $1.0 < a < 2.7$, for the case of interest, we can recast Eq. 2 into a simple relation between $a$ and $\theta$:

$$a(\theta) = 1 + \tfrac{17}{10}\sin\theta, \tag{3}$$

having used Eq. 1, and the expressions given above for $h_f$ and $\omega_P$. From Eq. 3 we can check again our limiting cases: $a(\theta = 0) = 1$, i.e., when the ball is released from $h_i = R$, will simply rise to the same final height ($h_f = R$, for $\theta = 0$); in the other extreme case $a(\theta = 90°) = 2.7$, when the ball is released from $h_i = 2.7R$ will "lose contact" at the top of the loop ($\theta = 90°$), but will still go around the loop.

**The Projectile Motion**

According to the analysis given above, if we release the ball from $h_i = aR$, with $1.0 < a < 2.7$, the moving object will lose contact with the loop track and become a "projectile," at an angular position, obtained from Eq. 3,

$$\theta = \sin^{-1}\left[\tfrac{10}{17}(a-1)\right]. \tag{4}$$

The ball will leave the track, with an "initial" velocity, as a projectile, which follows from Eq. 1:

$$v_P = \sqrt{gR\sin\theta} = \sqrt{\tfrac{10}{17}(a-1)gR}, \tag{5}$$

with the direction of this velocity vector described by the angle $\alpha$ formed with the horizontal,

$$\alpha = 90° - \theta. \tag{6}$$

For example, in Fig.1 we illustrate the case of the ball loosing contact at an angle $\theta = 30°$, which requires $a = \frac{37}{20} = 1.85$, or an initial release level $h_i = 1.85 R$. The ball becomes a projectile, "launched" with a speed $v_P = \sqrt{\frac{1}{2} gR}$, at angle $\alpha = 60°$ with the horizontal. In the same figure we also sketch the parabolic trajectory followed by the ball while falling inside the loop.[4]

An interesting position along the trajectory of this motion is represented by point H in Fig.1. This is the point at which the projectile is at the same level of the original launch position P. The horizontal distance $PH$ can be evaluated, by using the well-known (but often misused by students) formula for the horizontal range of a projectile.[5] Consider now the simple geometrical construction[6] in Fig. 1, where the line PA is tangent to the circle at P, point P' is at the same level of point P, segment P'A is drawn perpendicular to the tangent line and the vertical red line from A is intercepting the red horizontal line at H. Using Eqs. 5 and 6 (from which $\sin \alpha = \cos \theta$, $\cos \alpha = \sin \theta$) we obtain:

$$PH = \frac{v_P^2}{g} \sin 2\alpha = 2R \sin \theta \sin \alpha \cos \alpha = (2R \sin \alpha) \cos^2 \alpha =$$
$$= PP' \cos^2 \alpha = (PP' \cos \alpha) \cos \alpha = PA \cos \alpha. \qquad (7)$$

The previous equation, which applies for any possible value of the angle $\alpha$ (not just the case shown in figure), shows that point H can be found in general using the geometrical construction described above. In other words, to find H we simply draw the tangent to the circle at point P, then construct the perpendicular P'A from P' to the tangent line, where P' is a point on the circle at the same height of P. Point H will be the vertical projection of point A on the segment PP', as illustrated in Fig. 1 and mathematically proven in Eq. 7.

**A Modified Loop-the-Loop Experiment**

The discussion presented above suggests a different use of the loop-the-loop apparatus (see Fig. 2), in which we follow and analyze the projectile trajectory. To test the theory we filmed the motion of the *projectile-inside-the-loop* with a digital video camera and video capture software (VideoPoint 2.5), and then used video editing software to produce a "stroboscopic" picture of the motion. Video clips and photos of the experiment can be viewed on a related web page.[7]

Fig. 2 reproduces precisely the effect outlined in Fig. 1. The ball is launched from an initial release level $h_i = 1.85 R$, corresponding to an angle $\theta = 30°$ or $\alpha = 60°$, at which the ball is losing contact with the track. In this figure we obviously consider the motion of the *center-of-mass* of the sphere, so that the trajectory of the body is given at first by the black circle of radius R, somewhat smaller than the radius of the actual track. Apart from this difference the two figures represent essentially the same situation.

The ball then loses contact approximately at point P (for $\theta = 30°$) and proceeds along the parabolic path in figure. The center-of-mass positions are indicated by the red dots in the same figure. Using equation (7) the horizontal distance PH is computed as $PH = PP'\cos^2\alpha = \frac{1}{4}PP' = \frac{\sqrt{3}}{4}R$. This can actually be seen in Fig. 2, where the path of the falling ball, given by the red dots, appears to be going through point H, which is the intersection between the horizontal segment PP' and the vertical line from point A. In this particular case ($\theta = 30°$) the ball will also hit the track approximately at the central bottom point, as predicted above.

The use of the software allows for a better determination of the parabolic trajectory and all the related physical quantities. In particular, another interesting position is the top point of the parabolic trajectory. This corresponds to a height $h_{TOP} = a'R$, where $a'$ is a number between 1 and 2. In general $h_{TOP} < h_i$ (i.e., $a' < a$) because the sphere keeps rotating after losing contact with the track, with the same angular velocity it possessed at point P, therefore it cannot regain the same amount of potential energy it had at the beginning. Either by using conservation of mechanical energy, or by studying the parabolic trajectory, the relation between $a$ and $a'$ is

$$a' = a - \tfrac{1}{2}\cos^3\alpha - \tfrac{1}{5}\cos\alpha. \tag{8}$$

For the case of Fig. 2, with the initial release level at $h_i = 1.85R$, it is easy to calculate a value of $h_{TOP} = \frac{27}{16}R \cong 1.69R$, using the previous equation. Inspection of Fig. 2, done with the video software, shows complete agreement (the estimated position from the figure gives $h_{TOP} = 1.7R$).

We have repeated this experiment for different angles, such as $\theta = 45°, 60°$ and others, obtaining similar results.[8] The general analysis for the *projectile-inside-the-loop* and the geometrical considerations presented above are valid in any case and can be easily tested.

**Conclusion**

The *loop-the-loop* apparatus can be used also to study the kinematics of the projectile motion in a rather unusual way. This *projectile-inside-the-loop* experiment can be effectively integrated into classroom demonstrations or become part of laboratory activities. It can be used to show interesting connections between topics such as rolling motion, conservation of mechanical energy and projectile motion.

**Acknowledgments**

This research was supported by an award from Research Corporation. The author would like to acknowledge his friend and former teacher Prof. G. Tonzig, whose work inspired the original idea for this paper.

## References

1. PIRA demonstration 1M40.20, "loop the loop" and references therein. See PIRA web page at http://www.pira.nu/.
2. Several alternative uses of the classical loop-the-loop demonstration have been proposed. See for example PIRA demonstrations 1M40.21, 1M40.23, 1M40.24 and references therein (http://www.pira.nu/).
3. We recall that in a rolling motion of a rigid body without slipping, despite the presence of friction between the body and the track, no loss of mechanical energy occurs. This is due to the fact that the contact point is at rest relative to the surface at any instant.
4. In the particular case of $\theta = 30°$, shown in Fig. 1, it is also easy to prove that the ball, after falling, will strike the track exactly at the bottom point of the loop. We leave the proof to the reader. This of course will not happen for other values of the angle $\theta$.
5. The range equation for a projectile motion is usually reported as $R = v^2 \sin 2\alpha / g$, where $v$ and $\alpha$ are the initial speed and angle of the projectile, but this equation can be used to find the horizontal distance traveled only if the initial and final points considered are at the same height.
6. G. Tonzig, *Cento Errori di Fisica*, 1st ed. (Sansoni Editore, Firenze, Italy, 1991), p. 65.
7. The Projectile inside the Loop Web Page (http://myweb.lmu.edu/gvarieschi/loop/loop.html). This site contains all our photos and videos, plus links to other related pages.
8. For these particular angles the relevant quantities are as follows. For $\theta = 45°$, we have $h_i \cong 2.20\,R$, $PH = \frac{1}{2} PP' = \frac{\sqrt{2}}{2} R$, $h_{TOP} \cong 1.88\,R$. For $\theta = 60°$, we have $h_i \cong 2.47\,R$, $PH = \frac{3}{4} PP' = \frac{3}{4} R$, $h_{TOP} \cong 1.97\,R$.

**Gabriele U. Varieschi** is an assistant professor in the physics department at Loyola Marymount University. He earned his Ph.D. in theoretical particle physics from the University of California at Los Angeles. His research interests are in the area of astro-particle physics and cosmology.

**Department of Physics, Loyola Marymount University, 1 LMU Drive, Los Angeles, CA 90045; gvarieschi@lmu.edu**

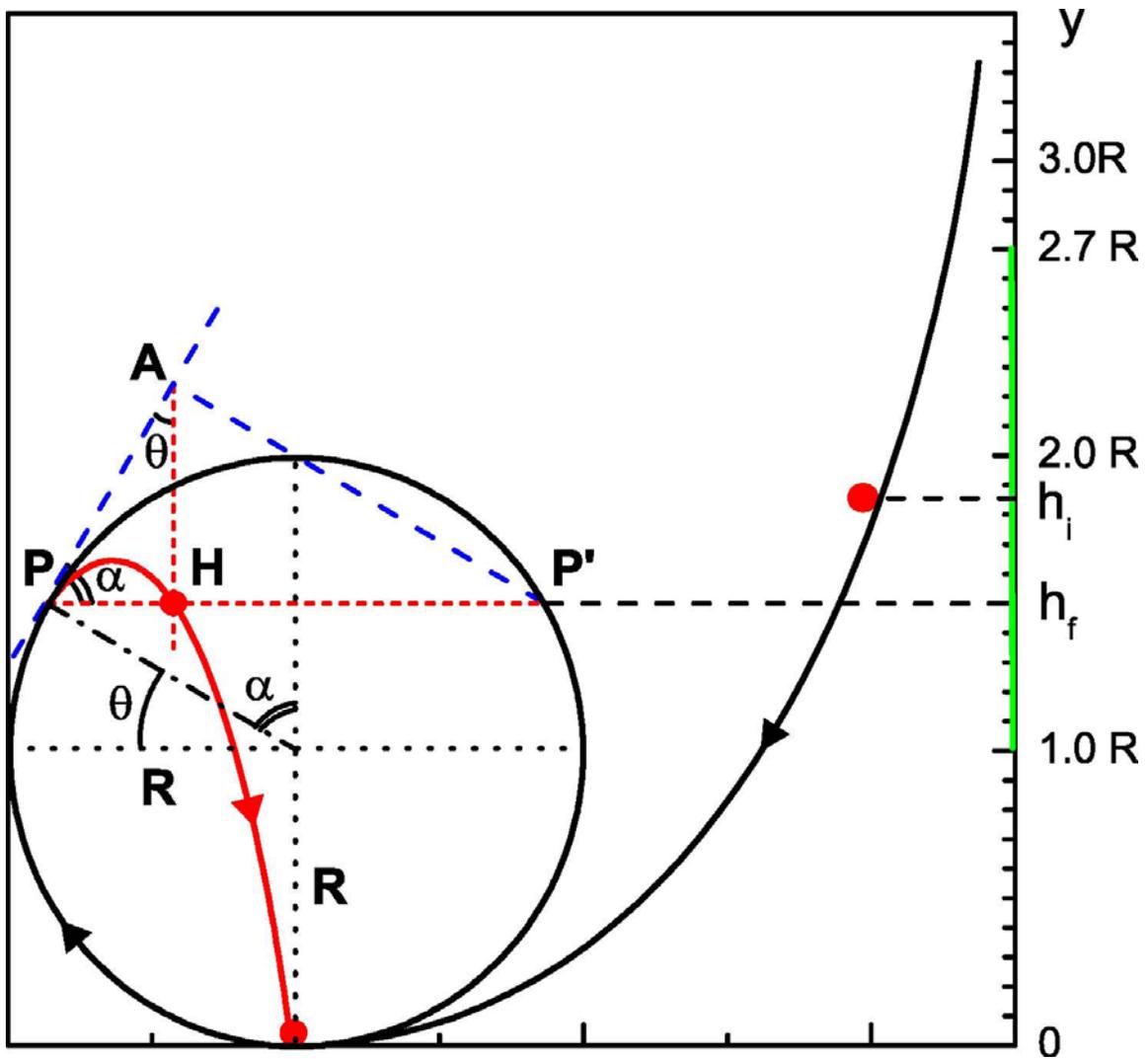

Fig. 1. Schematic description of the projectile inside the loop apparatus. For any initial position of the rolling sphere within the green vertical range, the moving body will lose contact with the circular track at position P and then will fall according to the parabolic trajectory, drawn in red. The geometrical construction used to determine point H is also shown for the particular case $\theta = 30°$.

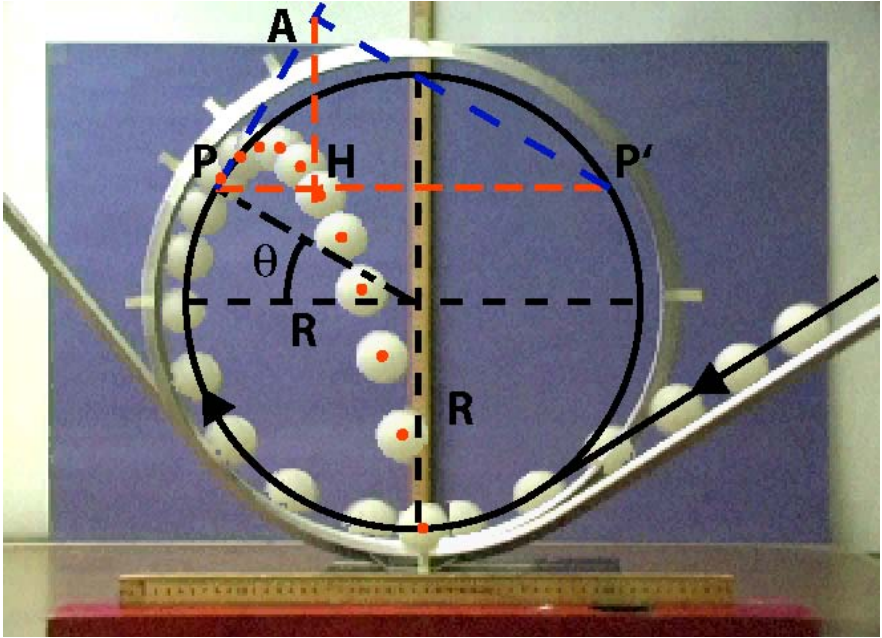

Fig. 2. Results of our demonstration for $\theta = 30°$. Video editing software was used to produce a "stroboscopic" picture of the motion. The position of the center-of-mass of the falling ball is indicated by the red dots. The geometrical construction is similar to the one in Fig. 1.